\documentclass[12pt]{article}
\usepackage{latexsym}
\usepackage{amsmath,amsfonts}
\usepackage{times}
\allowdisplaybreaks[4]
\catcode`\@=11

\newcount\hour
\newcount\minute
\newtoks\amorpm \hour=\time\divide\hour by 60\minute
=\time{\multiply\hour by 60 \global\advance\minute by-\hour}
\edef\standardtime{{\ifnum\hour<12 \global\amorpm={am}%
        \else\global\amorpm={pm}\advance\hour by-12 \fi
        \ifnum\hour=0 \hour=12 \fi
        \number\hour:\ifnum\minute<10
        0\fi\number\minute\the\amorpm}}
\edef\militarytime{\number\hour:\ifnum\minute<10 0\fi\number\minute}

\def\draftlabel#1{{\@bsphack\if@filesw {\let\thepage\relax
   \xdef\@gtempa{\write\@auxout{\string
      \newlabel{#1}{{\@currentlabel}{\thepage}}}}}\@gtempa
   \if@nobreak \ifvmode\nobreak\fi\fi\fi\@esphack}
        \gdef\@eqnlabel{#1}}
\def\@eqnlabel{}
\def\@vacuum{}
\def\marginnote#1{}
\def\draftmarginnote#1{\marginpar{\raggedright\scriptsize\tt#1}}
\def\draft{
        \pagestyle{plain}
        \overfullrule=2pt
        \oddsidemargin -.5truein
        \def\@oddhead{\sl \phantom{\today\quad\militarytime} \hfil
        \smash{\Large\sl DRAFT} \hfil \today\quad\militarytime}
        \let\@evenhead\@oddhead
        \let\label=\draftlabel
        \let\marginnote=\draftmarginnote
        \def\ps@empty{\let\@mkboth\@gobbletwo
        \def\@oddfoot{\hfil \smash{\Large\sl DRAFT} \hfil}
        \let\@evenfoot\@oddhead}
        \def\@eqnnum{(\theequation)\rlap{\kern\marginparsep\tt\@eqnlabel}%
        \global\let\@eqnlabel\@vacuum}  }

\newcommand{\rf}[1]{(\ref{#1})}
\renewcommand{\theequation}{\thesection.\arabic{equation}}
\renewcommand{\thefootnote}{\fnsymbol{footnote}}
\newcommand{\newsection}{   % Numeration of eqs. is automatic
\setcounter{equation}{0}\section}

\def\appendix#1{\addtocounter{section}{1}\setcounter{equation}{0}
\renewcommand{\thesection}{\Alph{section}}
\section*{Appendix \thesection\protect\indent \parbox[t]{11.15cm}{#1}}
\addcontentsline{toc}{section}{Appendix \thesection\ \ \ #1}}

\def\be{\begin{equation}}
\def\ee{\end{equation}}
\def\beq{\begin{eqnarray}}
\def\eeq{\end{eqnarray}}

\def\parline{\,\partial\kern -0.55em /\,\,}

\def\half{{\frac{1}{2}}}

\def\CC{{\cal C}}

\def\NN{{\cal N}}

\def\Rsm{{\scriptscriptstyle R}}
\def\Lsm{{\scriptscriptstyle L}}

\def\noinbf#1{\noindent {\bf #1}}

\def\mubf{{\boldsymbol{\mu}}}

\def\phik{|\phi\rangle}
\def\phibr{\langle\phi|}

\def\psik{|\psi\rangle}
\def\psibr{\langle\psi|}

\def\opsm{{\scriptscriptstyle \oplus}}
\def\omsm{{\scriptscriptstyle \ominus}}

\def\sm(A)dS{{\scriptscriptstyle (A)dS }}

\def\vph5{{\vphantom{5pt}}}

\def\Wb{\bar{W}}
\def\fb{\bar{f}}

\def\ub{\bar{u}}
\def\vb{\bar{v}}

\def\alphab{\bar\alpha}

\def\irm{{\rm i}}

\def\msv{{\rm msv}}
\def\pms{{\rm pms}}
\def\diff{{\rm diff}}
\def\field{{\rm field}}
\def\bfrm{{\rm bf}}

\def\for{{\rm for}}

\newcommand{\No}{\mathbb{N}}
\newcommand{\Zo}{\mathbb{Z}}

\begin{document}

%\draft

\begin{flushright}
FIAN-TD-2022-17  \hspace{1.9cm}  \ \\
arXiv: 2212.14728 V2 [hep-th] \\
\end{flushright}

\vspace{1cm}

\begin{center}

{\Large \bf Light-cone gauge massive and partially-massless

\medskip
fields in AdS(4)}

\vspace{2.5cm}

R.R. Metsaev%
\footnote{ E-mail: metsaev@lpi.ru
}

\vspace{1cm}

{\it Department of Theoretical Physics, P.N. Lebedev Physical
Institute, \\ Leninsky prospect 53,  Moscow 119991, Russia }

\vspace{2.5cm}

{}~   \hfill {\it To the memory of Lars Brink}

\vspace{2cm}

{\bf Abstract}

\end{center}

Using light-cone gauge approach, bosonic and fermionic massive and partially-massless fields in AdS(4) space are considered. For such fields, light-cone gauge action is presented. Considering the massive and partially-massless fields in helicity basis and CFT adapted basis, two simple representations of spin operators entering the light-cone gauge action are discussed. The simple representations for the spin operators are obtained by using bosonic spinor-like oscillators.
The bosonic spinor-like oscillators allow also us to treat the bosonic and fermionic fields on an equal footing.

\vspace{2cm}


\newpage
\renewcommand{\thefootnote}{\arabic{footnote}}
\setcounter{footnote}{0}

%%%%%%%%%%%%%%%%%%%%%%%%%%%%%%%%%%%%%%%%%%%
\newsection{\large Introduction}
%%%%%%%%%%%%%%%%%%%%%%%%%%%%%%

Light-cone gauge approach turns out to be powerful method for a study of field and string theories (for brief review see Ref.\cite{Brink:2005wh}). Among attractive applications of light-cone gauge approach one can mention the light-cone superspace formulation of closed superstring theories in Ref.\cite{Green:1983hw} and the proof of UV finiteness of the $\NN=4$ supersymmetric light-cone gauge Yang-Mills theory in Refs.\cite{Mandelstam:1982cb,Brink:1982wv}. Also it is worth mentioning the application  of light-cone gauge approach to the problem of interacting higher-spin massless fields in Refs.\cite{Bengtsson:1983pd,Bengtsson:1983pg}. Application of light-cone gauge for the study of counterterms  in gravity may be found in Ref.\cite{Bengtsson:2012dw}.

Light-cone gauge approach considerably simplifies Lagrangian formulation of fields in flat space. This triggered our interest in the generalization of light-cone approach to field dynamics in AdS space in Ref.\cite{Metsaev:1999ui}. In Ref.\cite{Metsaev:2003cu}, we applied the light-cone gauge approach for the study of bosonic and fermionic arbitrary spin massive fields in $AdS_{d+1}$, $d\geq 3$. In Ref.\cite{Metsaev:2003cu}, for the discussion of light-cone gauge massive fields, we used vector-like oscillators. However as is well-known, for the light-cone gauge description of massive fields in $AdS_4$, one can use bosonic spinor-like oscillators. It turns out that the use of the bosonic spinor-like oscillators simplifies considerably Lagrangian formulation of the light-cone gauge  massive fields in $AdS_4$. This is what we do in this paper. As a by product we obtain also light-cone gauge Lagrangian formulation of partially-massless fields.

Before to proceed let us mention some alternative approaches to massive and partially-massless fields. For the first time gauge invariant Lagrangian formulations of bosonic and fermionic massive and partially-massless fields were obtained by using the metric-like approach in the respective Refs.\cite{Zinoviev:2001dt} and \cite{Metsaev:2006zy}. Frame-like gauge invariant formulation for massive and partially-massless fields was developed in Refs.\cite{Zinoviev:2008ze}-\cite{Khabarov:2019dvi}. Lagrangian BRST formulation of AdS massive  fields was obtained in Refs.\cite{Buchbinder:2006ge}, while BRST equations of motion for partially-massless fields were discussed in Refs.\cite{Alkalaev:2009vm}.
For the first time, partial masslessness was discussed in pioneering reference \cite{Deser:1983mm}. The important conjecture about the masses for arbitrary spin partially-massless field in $AdS_4$ was made in Ref.\cite{Deser:2001xr}. This conjecture was proved in Refs.\cite{Zinoviev:2001dt,Metsaev:2006zy}. Also in Refs.\cite{Zinoviev:2001dt,Metsaev:2006zy} the masses of partially-massless fields in $AdS_{d+1}$ with arbitrary $d > 3$ were found.
For the updated list of references on partially-massless fields, see Ref.\cite{Basile:2022mif}.

\newsection{ \large Light-cone gauge action and relativistic symmetries of fields in $AdS_4$}

Light-cone gauge formulation in $AdS$ space was developed in Refs.\cite{Metsaev:1999ui,Metsaev:2003cu} (see also Refs.\cite{Metsaev:2019opn}). Here we adapt the formulation  in Refs.\cite{Metsaev:2003cu,Metsaev:2019opn} for the study of massive and partially-massless fields.

\noinbf{Light-cone gauge action}. To discuss light-cone gauge formulation of arbitrary spin field we introduce a ket-vector $\phik=|\phi(x,z,u,v)\rangle$. The arguments $x$, $z$, where $x\equiv x^+,x^-,x^1$ are used for the coordinates of $AdS_4$ space,%
\footnote{ Metric of $AdS_4$ space is given by, $ds^2=R^2(-dx^0dx^0+dx^1 dx^1 +dx^2dx^2+dzdz)/z^2$. We use the coordinates $x^\pm$ defined as $x^\pm=(x^2 \pm x^0)/\sqrt{2}$, where
$x^+$ is a light-cone evolution parameter. Our conventions for the derivatives are as follows: $\partial^1=\partial_1\equiv\partial/\partial x^1$,
$\partial_z\equiv\partial/\partial z$, $\partial^\pm=\partial_\mp
\equiv \partial/\partial x^\mp$.}
while the arguments $u$, $v$ stand for creation operators.
For integer spin-$s$ field, $s\in \No_0$, the ket-vector $\phik$ describes a collection of bosonic fields, while
for half-integer spin-$s$ field, $s\in \No_0+\half$, the ket-vector $\phik$ describes a collection of fermionic fields.
Ordinary light-cone gauge AdS fields which depend on the space-time coordinates $x$, $z$ are
obtainable by expanding the ket-vector $|\phi(x,z,u,v)\rangle$ into $u$ and $v$. Explicit expansions of the $\phik$ into $u$ and $v$ will be given below.
Light-cone gauge actions for the bosonic and fermionic fields in $AdS_4$ can be presented as
\beq
\label{05122022-manus-01} && S  = \half\int d^3 x dz\langle \phi|\bigl(\Box + \partial_z^2 - \frac{1}{z^2}A\bigr)|\phi\rangle\,, \hspace{1.7cm} \hbox{for bosonic field},
\nonumber\\
&& S  = \half\int d^3x dz \langle \phi|\frac{\irm }{\partial^+}\bigl(\Box + \partial_z^2 - \frac{1}{z^2}A\bigr)|\phi\rangle\,, \hspace{1.1cm} \hbox{for fermionic field},
\\
\label{05122022-manus-03}  && \hspace{1cm} \Box = 2\partial^+\partial^- + \partial^1\partial^1 \,,\qquad d^3x = dx^+dx^- dx^1\,,
\eeq
where an operator $A$ appearing in \rf{05122022-manus-01} depends only on the oscillators.  For massless arbitrary spin fields in $AdS_4$, we note the equality $A=0$ (see Ref.\cite{Metsaev:1999ui}).

\noinbf{Relativistic symmetries of fields in $AdS_4$}. Relativistic symmetries of fields propagating in $AdS_4$ are described by the $so(3,2)$ algebra.  The use of light-cone gauge spoils manifest symmetries of the $so(3,2)$ algebra. In order to show that symmetries of the $so(3,2)$ algebra are maintained we should to present the Noether charges (generators) which generate them. For free bosonic and fermionic fields, Noether generators take the following respective forms:
\beq
\label{05122022-manus-04} &&    G_\field =  \int dz dx^- dx^1 \, \langle\partial^+\phi|G_\diff|\phi\rangle\,, \hspace{2cm} \hbox{for bosonic field},
\nonumber\\
&& G_\field =  -\irm \int dz dx^- dx^1\, \langle \phi|G_\diff|\phi\rangle\,, \hspace{2cm} \hbox{for fermionic field},
\eeq
where $G_\diff$ stand for differential operators acting on the ket-vectors.  Actions  \rf{05122022-manus-01} are invariant under the transformations $\delta\phik =G_\diff\phik$, where the operators $G_\diff$ are given by%
\footnote{ Our conventions for commutators of the generators \rf{05122022-manus-06}-\rf{05122022-manus-14} may be found in (A6),(A7) in Ref.\cite{Metsaev:2015rda}.
}
\beq
\label{05122022-manus-06} && P^1=\partial^1\,, \qquad  P^+=\partial^+\,,\qquad P^-=-\frac{\partial^1\partial^1+\partial_z\partial_z}{2\partial^+} +\frac{1}{2z^2\partial^+}A\,,
\\
\label{05122022-manus-07} && J^{+1}=x^+\partial^1-x^1\partial^+\,,\hspace{1.3cm} J^{-1}=x_\bfrm^-\partial^1-x^1 P^- +M^{-1}\,,
\\
\label{05122022-manus-08} && J^{+-}=x^+ P^- -x_\bfrm^-\partial^+\,,
\\
\label{05122022-manus-09} && D=x^+ P^- +x_\bfrm^-\partial^+ + x^1\partial^1+ z\partial_z + 1\,,
\\
\label{05122022-manus-10} && K^+ = -\frac{1}{2}(2x^+x_\bfrm^-+ z^2 + x^1 x^1)\partial^+ + x^+D\,,
\\
\label{05122022-manus-11} && K^1 = - \half (2x^+x_\bfrm^-+ z^2 + x^1x^1)\partial^1  + x^1 D - M^{z1} z - M^{-1} x^+\,,
\\
\label{05122022-manus-12} && K^- = - \half (2x^+x_\bfrm^- + z^2 + x^1 x^1) P^- + x_\bfrm^-D
\nonumber\\
&& \hspace{0.9cm} +\,\, \frac{1}{\partial^+}(z\partial^1-x^1\partial_z)M^{z1}
- \frac{x^1}{2z\partial^+}[M^{z1},A] + \frac{1}{\partial^+}B\,,\qquad
\\
\label{05122022-manus-14} && \hspace{0.8cm} M^{-1} =  - M^{z1}\frac{\partial_z}{\partial^+}
-\frac{1}{2z\partial^+}[M^{z1},A]\,,
\\
&& \hspace{0.8cm} x_{\bfrm}^- = x^- \,,  \hspace{0.2cm} \hbox{for bosonic field}; \hspace{0.8cm}  x_{\bfrm}^- = x^- + \frac{1}{2\partial^+}\,, \hspace{0.2cm} \hbox{for fermionic field}.\qquad
\eeq

Operators $A$, $B$, $M^{z1}$ appearing in \rf{05122022-manus-06}-\rf{05122022-manus-14} depend only on the oscillators. The operator $B$ is expressed in terms of the operator $A$. The operator $A$ is AdS cousin of a flat space mass operator, while the operator $M^{z1}$ is AdS cousin of a flat space helicity operator.  For massive and partially-massless fields, the operators $A$ and $M^{z1}$ are not commuting when $s\ne 0$. Therefore  they cannot be diagonalized simultaneously. This motivates us to introduce two bases for AdS fields which we refer to as helicity basis and CFT adapted basis. These two bases are defined as
\be \label{05122022-manus-16}
M^{z1} \hbox{ is diagonal for helicity basis}; \hspace{1cm} A \hbox{ is diagonal for CFT adapted basis}.
\ee
As the basis with the diagonalized operator $A$ turned out to be convenient for the study of light-cone gauge AdS/CFT correspondence in Ref.\cite{Metsaev:2015rda} we refer to such basis as CFT adapted basis.

The bra-vector $\phibr$ in \rf{05122022-manus-01}, \rf{05122022-manus-04} and hermitian properties of the operators $A$, $M^{z1}$, $B$ are  given by
\be \label{05122022-manus-17a1}
\phibr \equiv \phik^\dagger\mubf\,, \quad \ (\mubf A)^\dagger = \mubf A\,, \quad \ (\mubf M^{z1})^\dagger = - \mubf M^{z1}\,, \quad \ (\mubf B)^\dagger = \mubf B\,,
\ee
where an operator $\mubf$ depends only on the oscillators and satisfies the relations $\mubf^\dagger=\mubf$, $\mubf^2=1$. For massive field we use $\mubf=1$, while, for partially-massless field, the $\mubf$ will be given below.

Throughout this paper we use bosonic spinor-like creation operators $u$, $v$ and the respective annihilation operators $\ub$, $\vb$ which we refer to as oscillators,
\be \label{05122022-manus-18}
[\ub,u]=1\,, \qquad [\vb,v]=1\,, \qquad \ub|0\rangle = 0\,, \qquad \vb|0\rangle = 0\,, \qquad  u^\dagger = \ub\,, \qquad v^\dagger = \vb\,.
\ee
Generators of the $so(3)$ algebra $S^\Rsm$, $S^\Lsm$, and $S$ and ket-vectors $|n\rangle$ are defined by the relations
\beq
\label{05122022-manus-19} && S^\Rsm  = \frac{1}{\sqrt{2}}u\vb \,, \qquad \ S^\Lsm  =  \frac{1}{\sqrt{2}} v\ub \,,\qquad \ \ S  = \half ( u\ub - v\vb)\,,
\nonumber\\
\label{05122022-manus-20} && [S,S^\Rsm ] =S^\Rsm\,, \qquad [S,S^\Lsm ] = - S^\Lsm\,, \qquad  [S^\Rsm  , S^\Lsm ] = S \,,
\\
\label{05122022-manus-21} && \hspace{2cm} |n\rangle \equiv \frac{u^{s+n} v^{s-n}}{\sqrt{(s+n)!(s-n)!}} |0 \rangle\,.
\eeq
For two integers $a,b\in \Zo $ (or half-integers $a,b\in \Zo+\half$), $a\leq b$, we use the convention
\be
n\in [a,b] \quad \Longleftrightarrow \quad n=a,a+1,\ldots, b-1,b\,.
\ee

%%%%%%%%%%%%%%%%%%%%%%%%%%%%%%%%%%%%%%%%%%%%%%%%%%%%%%%%%%%%%%%%%%%%%%%%%%%%%%%%%%%%%%%%%
%%%%%%%%%%%%%%%%%%%%%%%%%%%%%%%%%%%%%%%%%%%%%%%%%%%%%%%%%%%%%%%%%%%%%%%%%%%%%%%%%%%%%%%%%
\newsection{ \large Massive and partially-massless fields in helicity basis } \label{sec-03}
%%%%%%%%%%%%%%%%%%%%%%%%%%%%%%%%%%%%%%%%%%%%%%%%%%%%%%%%%%%%%%%%%%%%%%%%%%%%%%%%%%%%%%%%%
%%%%%%%%%%%%%%%%%%%%%%%%%%%%%%%%%%%%%%%%%%%%%%%%%%%%%%%%%%%%%%%%%%%%%%%%%%%%%%%%%%%%%%%%%

\noinbf{ Massive fields in helicity basis}. For light-cone gauge description of spin-$s$ massive field we use ket-vector $\phik$ given by
\be \label{06122022-manus-01}
\phik = \sum_{n\in [-s,s]} \phi_n(x,z) |n\rangle\,,\qquad (\phi_n(x,z))^\dagger = \phi_{-n}(x,z)\,,
\ee
where, for spin-$s$ bosonic field, $s\in \No_0$, while, for fermionic spin-$s$ massive field, $s\in \No_0+\half$. We should provide a realization of the operators $A$, $B$ and $M^{z1}$ on  ket-vector \rf{06122022-manus-01}. Here we present our result for the operators $A$, $M^{z1}$, and $B$,
\beq
&&  A = \CC_2  + 2 -2S^2 + \irm \sqrt{2}(S^\Lsm  \fb - f S^\Rsm)   \,, \hspace{1cm} M^{z1} = - \irm S\,,
\nonumber\\
\label{06122022-manus-07} && B = - S^2 +  \frac{\irm}{\sqrt{2}}(S^\Lsm  \fb - f S^\Rsm)  \,,\qquad
\\
\label{06122022-manus-08} && f = F\,, \qquad \fb  = F\,, \qquad F \equiv \big((E_0  -1 - S)(E_0  - 2 + S)\big)^\half\,,
\\
\label{06122022-manus-09} && \CC_2  \equiv E_0(E_0-3) + s(s+1)\,,
\eeq
where $S^{\Rsm,\Lsm}$, $S$ are given in \rf{05122022-manus-19}. An energy parameter $E_0$ \rf{06122022-manus-09} stands for lowest eigenvalue of the energy operator of the $so(3,2)$ algebra  irrep  that is associated with the massive spin-$s$ field, while $\CC_2$ \rf{06122022-manus-09} stands for the corresponding eigenvalue of the 2nd order Casimir operator. For the brief derivation of \rf{06122022-manus-07}, \rf{06122022-manus-08}, see Sec.\ref{sec-05}. The following remarks are in order.

\noinbf{i})  For scalar field ($s=0$), spin-$s$ bosonic field ($s\in \No$), and spin-$s$ fermionic field ($s\in \No_0+\half$), the $E_0$ is expressed in terms of mass parameter $m$ by the well known relations (see, e.g. Refs.\cite{Metsaev:2003cu,Metsaev:2006zy})
\beq
\label{06122022-manus-10} && E_0\big|_{s=0} = \frac{3}{2} \pm  \Big(m^2 + \frac{9}{4}\Big)^\half, \qquad
 E_0\big|_{s\in \No}  = \frac{3}{2} + \Big(m^2 + \big(s-\half\big)^2\Big)^\half\,,
\nonumber\\
\label{06122022-manus-11} && E_0\big|_{s\in \No_0+\half}  = \frac{3}{2} + m\,.
\eeq
Accordingly, in terms of the mass parameter $m$, the respective values of $\CC_2$ \rf{06122022-manus-09} take the form
\beq
\label{06122022-manus-12} && \CC_2|_{s=0} = m^2\,, \hspace{1cm} \CC_2|_{s\in \No} = m^2 + 2s^2 -2\,,
\nonumber\\
&& \CC_2|_{s\in \No_0+\half} = m^2 + s(s+1) - \frac{9}{4}\,.
\eeq
Comparing \rf{06122022-manus-09} and \rf{06122022-manus-12}, we see that it is the use of the $E_0$ that provides us the universal expression for $\CC_2$ \rf{06122022-manus-09} which is valid for arbitrary spin massive fields.

\noinbf{ii}) We recall that, in the massless limit, the mass parameters and $\CC_2$ are given by
\beq
\label{06122022-manus-14} && m^2|_{s=0}=-2\,, \hspace{1cm} m|_{s\in \No}=0\,,  \qquad m|_{s\in \No_0+\half} =s-\half\,,
\nonumber\\
&& \CC_2\big|_{s\in \No_0, \No_0+\half} = 2s^2-2\,,  \hspace{2.4cm} \hbox{for massless fields}.
\eeq

\noinbf{iii}) We recall the unitarity restriction in Ref.\cite{Evans},%
\footnote{Generalization of \rf{06122022-manus-15} to arbitrary spin mixed-symmetry fields  in $AdS_{d+1}$, $d>4$, may be found in Ref.\cite{Metsaev:1995re}. For brief review of the unitarity restrictions for some irreps of the $so(d,2)$ algebra, see Ref.\cite{Ponomarev:2022vjb}.}
\be \label{06122022-manus-15}
E_0> s+1\,, \hspace{5.7cm} \hbox{for spin-$s$ massive field}.
\ee
Using \rf{06122022-manus-15}, we verify that eigenvalues of the operator $F$ \rf{06122022-manus-08} on space of ket-vector $\phik$ \rf{06122022-manus-01} are real-valued. Therefore, in  \rf{05122022-manus-17a1}, we can use the simplest choice $\mubf=1$. With this choice, the operators $A$, $M^{z1}$, $B$ \rf{06122022-manus-07} satisfy the hermitian conjugation rules given in \rf{05122022-manus-17a1}.

\noinbf{iv}) In view of the relation $S|n\rangle=n|n\rangle$ (see \rf{05122022-manus-19} and  \rf{06122022-manus-01}), the operator $M^{z1}$ \rf{06122022-manus-07} is diagonal on space of the fields $\phi_n(x,z)$ \rf{06122022-manus-01}.

\noinbf{Partially-massless fields in helicity basis}. Consider irreps of the $so(3,2)$ algebra with the following values of the energy parameter $E_0$:
\be \label{06122022-manus-35}
E_0 = s + 1 - t\,, \qquad t=1,2,\ldots, [s]-1\,, \quad [s]> 1\,,\qquad  \hbox{for partially-massless field}.
\ee
Field in $AdS_4$ associated with the $so(3,2)$ algebra irrep having $E_0$ as in \rf{06122022-manus-35} is referred to spin-$s$ and depth-$t$ partially-massless field.
For $t=0$, we get $E_0$ corresponding to spin-$s$ massless field.%
\footnote{Our parameter $t$ is related to the commonly used depth of partially-massless field $t_{\rm com.us}$ as $t = t_{\rm com.us}-1$.}

For light-cone gauge description of spin-$s$ and depth-$t$ partially-massless field, we introduce the ket-vector $\phik$ defined as
\beq
\label{06122022-manus-36} && \hspace{-0.8cm} \phik = |\phi^\opsm\rangle + |\phi^\omsm\rangle\,, \qquad |\phi^\opsm\rangle = \sum_{n\in [s-t,s]} \phi_n(x,z) |n\rangle\,,\quad
|\phi^\omsm\rangle = \sum_{n\in [-s,-s+t]} \phi_n(x,z) |n\rangle\,,\qquad
\\
\label{06122022-manus-37} && \hspace{3.7cm} (\phi_n(x,z))^\dagger = \phi_{-n}(x,z)\,.
\eeq
Helicity of the component field $\phi_n(x,z)$ is equal to $n$. Therefore the ket-vectors $|\phi^\opsm\rangle$ and $|\phi^\omsm\rangle$ \rf{06122022-manus-36} describe the respective positive and negative helicity component fields related by the hermicity conjugation rule \rf{06122022-manus-37}. The ket-vector $\phik$ \rf{06122022-manus-36} consists of $2t+2$ component fields as it should be for depth-$t$ partially-massless field in $AdS_4$.%
\footnote{For massless fields, $t=0$. This implies that the partially-massless fields have more D.o.F than massless fields in the flat space. In this respect the partially-massless fields are similar to mixed-symmetry fields in $AdS_{d+1}$, $d\geq 5$ studied in Ref.\cite{Brink:2000ag}. Note however that the mixed-symmetry AdS fields are associated with unitary irreps of the $so(d,2)$ algebra, while the partially-massless fields are associated with non-unitary irreps of the $so(d,2)$ algebra.
}

Using $E_0$ \rf{06122022-manus-35}, we note that realization of the operators $A$, $M^{z1}$, and $B$ on ket-vectors $|\phi^{\opsm,\omsm}\rangle$ \rf{06122022-manus-36} takes the same form as in \rf{06122022-manus-07}, where operators $f$ and $\fb$ are given by
\beq
\label{06122022-manus-38} && f = e^{\irm \pi (S-s)} |F| \,, \qquad  \fb = e^{\irm \pi (S-s-1)} |F|\,, \hspace{1cm} \for \ \ |\phi^\opsm\rangle\,;
\nonumber\\
&& f = e^{\irm \pi (S+s)} |F| \,, \qquad  \fb = e^{\irm \pi (S+s-1)} |F|\,, \hspace{1cm} \for \ \ |\phi^\omsm\rangle\,;
\nonumber\\
&& \hspace{1.5cm} F = \big((s  - t - S)(s - 1 - t + S)\big)^\half\,,\qquad
\eeq
while, in the expression for $\CC_2$ \rf{06122022-manus-09}, we use $E_0$ given in \rf{06122022-manus-35}.
The bra-vector $\phibr$ entering action of partially-massless field \rf{05122022-manus-01} is expressed in terms of the ket-vectors $|\phi^{\opsm,\omsm}\rangle$ as
\beq
\label{06122022-manus-39} && \phibr \equiv \langle\phi^\opsm|\mubf^\opsm + \langle\phi^\omsm|\mubf^\omsm\,, \qquad \mubf^\opsm = e^{\irm \pi (S-s)}\,,\qquad \mubf^\omsm = e^{\irm \pi (S+s)}\,,
\eeq
where $\langle\phi^\opsm|\equiv |\phi^\opsm\rangle^\dagger$, $\langle\phi^\omsm|\equiv |\phi^\omsm\rangle^\dagger$. Using $\mubf^\opsm, \mubf^\omsm$ \rf{06122022-manus-39}, we verify that the operators $A$, $M^{z1}$, $B$ satisfy the hermitian conjugation rules given in \rf{05122022-manus-17a1}. For example $(\mubf^\opsm A)^\dagger = \mubf^\opsm A$, and so on.  Using the notation $S(\phi)$ for the action in \rf{05122022-manus-01}, we note that the action \rf{05122022-manus-01} is factorized as $S(\phi)= S(\phi^\opsm)+ S(\phi^\omsm)$.
The action $S(\phi^\opsm)$ is constructed in terms of the ket-vector $|\phi^\opsm\rangle$ and the bra-vector $\langle \phi^\opsm|$. The ket-vector $|\phi^\opsm\rangle$ \rf{06122022-manus-36} is decomposed into the fields with the positive helicities $\lambda\in [s-t,s]$, while the bra-vector $\langle\phi^\opsm|\equiv |\phi^\opsm\rangle^\dagger$, in view of \rf{06122022-manus-37}, is decomposed
into the fields with the negative helicities $\lambda\in [-s,-s+t]$. Therefore the action $S(\phi^\opsm)$ is built from the pairs of the fields of opposite helicities.
The same holds true for the action $S(\phi^\omsm)$ which is constructed in terms of the ket-vector $|\phi^\omsm\rangle$ and the bra-vector $\langle \phi^\omsm|$. The ket-vector $|\phi^\omsm\rangle$ \rf{06122022-manus-36} is decomposed into the fields with the negative helicities $\lambda\in [-s,-s+t]$, while the bra-vector $\langle\phi^\omsm|\equiv |\phi^\omsm\rangle^\dagger$, in view of \rf{06122022-manus-37}, is decomposed
into the fields with the positive helicities $\lambda\in [s-t,s]$. Therefore the action $S(\phi^\omsm)$ is also built from the pairs of the fields of opposite helicities.
Moreover, in view of \rf{06122022-manus-37}, one has the equality $S(\phi^\opsm)= S(\phi^\omsm)$.
Note that the action $S(\phi)= S(\phi^\opsm)+ S(\phi^\omsm)$ appears naturally upon consideration  of the partial-massless limit in the action for massive fields (see relations \rf{08122022-manus-16}, \rf{08122022-manus-17} in Sec.\ref{sec-05}.)

\noinbf{Partially-massless fields in helicity basis. Alternative formulation}.  We introduce the ket-vector $\psik$ related to the ket-vectors $|\phi^\opsm\rangle$, $|\phi^\omsm\rangle$ \rf{06122022-manus-36} as%
\footnote{ Our motivation for the alternative formulation is related to the fact that, as compared to the expansions of $|\phi^{\opsm,\omsm}\rangle$ in \rf{06122022-manus-36}, the ket-vectors $|\psi^{\opsm,\omsm}\rangle$ have more elegant expansions in the oscillators $u$ and $v$ \rf{06122022-manus-51}.
}
\beq
\label{06122022-manus-50} && \hspace{-1.5cm} \psik = |\psi^\opsm\rangle + |\psi^\omsm\rangle\,, \qquad  |\phi^\opsm\rangle = U_{u,2s-t}|\psi^\opsm\rangle\,, \qquad |\phi^\omsm\rangle = U_{v,2s-t}|\psi^\omsm\rangle\,,
\\
\label{06122022-manus-51} && |\psi^{\opsm,\omsm}\rangle = \sum_{n\in [-\frac{t}{2},\frac{t}{2}]} \frac{u^{\frac{t}{2}+n}v^{\frac{t}{2}-n}}{\sqrt{(\frac{t}{2}+n)!(\frac{t}{2}-n)!}} \psi_n^{\opsm,\omsm}(x,z)|0\rangle \,,
\\
\label{06122022-manus-52} && U_{\alpha,k} \equiv  \alpha^k \NN_{\alpha,k}\,,\qquad \NN_{\alpha,k} \equiv \big((\alpha\alphab+1)\ldots (\alpha\alphab+k)\big)^{-\half}\,,\qquad U_{\alpha,k}^\dagger U_{\alpha,k}^\vph5=1\,, \qquad
\\
\label{06122022-manus-53}
&& \psi_n^\opsm = \phi_{n+s - \frac{t}{2}}\,, \qquad \psi_n^\omsm = \phi_{n-s + \frac{t}{2}}\,, \qquad \psi_n^\opsm{}^\dagger = \psi_{-n}^\omsm\,,
\eeq
where \rf{06122022-manus-53} are obtained from  \rf{06122022-manus-50}-\rf{06122022-manus-52} and \rf{06122022-manus-37}. The operators $A$, $M^{z1}$, $B$ are found to be
\beq
\label{06122022-manus-54} &&  A = \CC_2  + 2 -2M^{\Rsm\Lsm} M^{\Rsm\Lsm} +  \irm \sqrt{2}(S^\Lsm  \fb - f S^\Rsm)  \,, \hspace{1cm} M^{z1} = - \irm M^{\Rsm\Lsm}\,,
\nonumber\\
&& B = - M^{\Rsm\Lsm} M^{\Rsm\Lsm} +  \frac{\irm}{\sqrt{2}} (S^\Lsm  \fb - f S^\Rsm)  \,,\qquad
\eeq
where the realization of the operators $M^{\Rsm\Lsm}$, $f$ and $\fb$ on the ket-vectors $|\psi^{\opsm,\omsm}\rangle$ is given by
\beq
\label{06122022-manus-55} && \hspace{-2cm} M^{\Rsm\Lsm} =  S + s - \half t\,, \qquad f =  e^{\irm\pi (S-\half t)} F\,, \qquad \fb = e^{\irm\pi (S-\half t - 1)} F\,,
\nonumber\\
&& \hspace{-2cm} F = \big((2s- \half t +S)(2s-1 -\frac{3}{2} t + S)\big)^\half\,, \hspace{4cm}
\for \quad |\psi^\opsm\rangle\,;
\\
\label{06122022-manus-56} && \hspace{-2cm} M^{\Rsm\Lsm} =  S - s + \half t\,, \qquad f =  e^{\irm\pi (S+\half t)} F\,, \qquad \fb = e^{\irm\pi (S+\half t - 1)} F\,,
\nonumber\\
&& \hspace{-2cm} F = \big((2s + 1 - \half t - S)(2s -\frac{3}{2} t - S)\big)^\half\,, \hspace{4cm} \for \quad |\psi^\omsm\rangle\,;
\eeq
while $\CC_2$ in \rf{06122022-manus-54} is obtained by using \rf{06122022-manus-09} and  \rf{06122022-manus-35}. The bra-vector $\psibr$ is defined as
\beq
\label{06122022-manus-57}  && \psibr \equiv \langle\psi^\opsm|\mubf^\opsm + \langle\psi^\omsm|\mubf^\omsm\,, \qquad
\mubf^\opsm = e^{\irm \pi (S-\frac{t}{2})}\,,\qquad \mubf^\omsm = e^{\irm \pi (S+\frac{t}{2})}\,,
\eeq
where $\langle\psi^\opsm| \equiv |\psi^\opsm\rangle^\dagger$, $\langle\psi^\omsm| \equiv |\psi^\omsm\rangle^\dagger$. Using $\mubf^\opsm, \mubf^\omsm$ \rf{06122022-manus-57}, we verify that the operators $A$, $M^{z1}$, $B$ satisfy the rules given in \rf{05122022-manus-17a1}. For example $(\mubf^\opsm A)^\dagger = \mubf^\opsm A$, and so on. Action for $\psik$ is obtained by using the replacement $\phi \rightarrow \psi$ in \rf{05122022-manus-01}. The action $S(\psi)$ is factorized as $S(\psi)= S(\psi^\opsm)+ S(\psi^\omsm)$. Moreover, in view of $(\psi_n^\opsm)^\dagger = \psi_{-n}^\omsm$ \rf{06122022-manus-53} one has the equality $S(\psi^\opsm)=S(\psi^\omsm)$.

%%%%%%%%%%%%%%%%%%%%%%%%%%%%%%%%%%%%%%%%%%%%%%%%%%%%%%%%%%%%%%%%%%%%%%%%%%%%%%%%%%%%%%%%%
%%%%%%%%%%%%%%%%%%%%%%%%%%%%%%%%%%%%%%%%%%%%%%%%%%%%%%%%%%%%%%%%%%%%%%%%%%%%%%%%%%%%%%%%%
\newsection{ \large Massive and partially-massless fields in CFT adapted basis } \label{sec-04}
%%%%%%%%%%%%%%%%%%%%%%%%%%%%%%%%%%%%%%%%%%%%%%%%%%%%%%%%%%%%%%%%%%%%%%%%%%%%%%%%%%%%%%%%%
%%%%%%%%%%%%%%%%%%%%%%%%%%%%%%%%%%%%%%%%%%%%%%%%%%%%%%%%%%%%%%%%%%%%%%%%%%%%%%%%%%%%%%%%%

\noinbf{Massive fields in CFT adapted basis}. For light-cone gauge description of massive bosonic and fermionic fields in CFT adapted basis we use ket-vector $\phik$ defined as
\be \label{07122022-manus-01}
\phik = \sum_{n\in [-s,s]} \phi_n(x,z) |n\rangle\,,\qquad (\phi_n(x,z))^\dagger = \phi_n(x,z)\,,
\ee
where all component fields are real-valued and $|n\rangle$ is defined in \rf{05122022-manus-21}.
Realization of the operators $A$, $M^{z1}$, and $B$ on the ket-vector $\phik$ \rf{07122022-manus-01} is given by
\beq
\label{07122022-manus-15} && A = \nu^2 - \frac{1}{4}\,, \hspace{2.3cm} \nu^\vph5 \equiv \kappa+ S \,, \hspace{1cm} \kappa  \equiv E_0 - \frac{3}{2}\,,
\nonumber\\
&& M^{z1} =  f  S^\Rsm - S^\Lsm  \fb \,, \hspace{1cm} B =  \kappa S + \half \big(S^2- s(s+1))\,, \hspace{1.5cm} %
\\
\label{07122022-manus-18} && f= F\,, \qquad \fb =F\,, \qquad F \equiv \Big(\frac{(2\kappa  + s+S)(2\kappa  -  s -1 + S)}{ 8(\kappa  + S)
(\kappa  -1 + S)}\Big)^\half\,,\qquad
\eeq
where $S^{\Rsm,\Lsm}$, $S$ are given in \rf{05122022-manus-19}, while the energy parameter $E_0$ \rf{07122022-manus-15} satisfies the unitary restriction \rf{06122022-manus-15}. In view of the relation $S|n\rangle=n|n\rangle$, the operator $A$ \rf{07122022-manus-15} is diagonal on $\phik$ \rf{07122022-manus-01}.
Using \rf{06122022-manus-15}, we verify that eigenvalues of the operator $F$ \rf{07122022-manus-18} on space of ket-vector  $\phik$ \rf{07122022-manus-01} are real-valued.
Therefore, in  \rf{05122022-manus-17a1}, we can use the simplest choice $\mubf=1$. With such choice the operators $A$, $M^{z1}$, $B$ \rf{07122022-manus-15} satisfy the hermitian conjugation rules given in \rf{05122022-manus-17a1}.

\noinbf{Partially-massless fields in CFT adapted basis}. For light-cone gauge description of partially-massless  bosonic and fermionic fields in CFT adapted basis we use ket-vector $\phik$ given by
\be \label{07122022-manus-19}
|\phi\rangle = \sum_{n\in [-s,-s+1+2t]} \phi_n(x,z) |n\rangle\,, \qquad (\phi_n(x,z))^\dagger = \phi_n(x,z))\,,
\ee
where component fields $\phi_n(x,z)$ are real-valued. Ket-vector $\phik$ \rf{07122022-manus-19} consists of $2t+2$ component fields as it should be for depth-$t$ partially-massless field in $AdS_4$. Using $E_0$ as in \rf{06122022-manus-35}, we note that
realization of the operators $A$, $M^{z1}$, and $B$ on ket-vector $\phik$ \rf{07122022-manus-19}  takes the same form as in \rf{07122022-manus-15}, where $f$ and $\fb$ are given by
\beq
\label{07122022-manus-20}  && f=e^{\irm \pi \rho_S^\vph5 } |F|\,, \qquad \fb=e^{\irm \pi \rho_{S-1}^\vph5 }|F|\,,\qquad \rho_S^\vph5 =  |S+s-t-\half| - t - \half\,,\qquad\qquad
\nonumber\\
&& F = \Big(\frac{(3s-1-2t+S)(s-2-2t+S)}{2(4(s-1-t+S)^2-1)}\Big)^\half\,.
\eeq
The bra-vector $\phibr$ entering action \rf{05122022-manus-01} is expressed in terms of ket-vector $\phik$ \rf{07122022-manus-19} as $\langle \phi| = \phik^\dagger \mubf$ where $\mubf  = e^{\irm \pi \rho_S^\vph5}$, while $\rho_S^\vph5$ is given in \rf{07122022-manus-20}. Note that $\rho_S^\vph5=|\nu|-t-\half$, where $\nu$ is given by \rf{07122022-manus-15} and \rf{06122022-manus-35}. We also recall the relation  $\mubf^2\phik=\phik$.

\noinbf{Partially-massless fields in CFT adapted basis. Alternative formulation}. We introduce ket-vector $\psik$ related to the ket-vector $\phik$ \rf{07122022-manus-19} as
\beq
\label{07122022-manus-27} && \hspace{-1cm} |\phi\rangle  = U_{v,2s-1-2t}  |\psi\rangle \,,
\\
\label{07122022-manus-28} && \hspace{-1cm} |\psi\rangle = \sum_{n\in [-t-\half,t+\half]} \frac{u^{t+\half+n}v^{t+\half-n}}{\sqrt{(t+\half+n)!(t+\half-n)!}} \psi_n(x,z)|0\rangle \,,\qquad \phi_{n-s+t+\half} = \psi_n\,,
\eeq
where the component fields $\psi_n(x,z)$ are real-valued, while the operator $U_{v,2s-1-2t}$ is defined as in \rf{06122022-manus-52}. The realization of the operators $A$, $M^{z1}$, $B$ on the $\phik$ given in \rf{07122022-manus-15}, \rf{07122022-manus-20} and the transformation \rf{07122022-manus-28} lead to  the following realization of the operators $A$, $M^{z1}$, $B$ on the $\psik$:
\beq
\label{07122022-manus-31} && A = S^2 - \frac{1}{4}\,,\qquad B = \half \big(S^2 - \CC_2 - \frac{9}{4} \big)\,, \qquad M^{z1} =  f  S^\Rsm - S^\Lsm  \fb \,,
\\
\label{07122022-manus-32}  && f=e^{\irm \pi \rho_S^\vph5} |F|\,, \qquad \fb=e^{\irm \pi \rho_{S-1}^\vph5}|F|\,,\qquad \rho_S^\vph5 =  |S| - t - \half\,,\qquad\qquad
\nonumber\\
\label{07122022-manus-33} && F = \Big(\frac{(2s-t -\half +S)(2s - t + \half -S)}{8S(S-1)}\Big)^\half\,,
\eeq
where $\CC_2$ is given in \rf{06122022-manus-09} and \rf{06122022-manus-35}. The bra-vector $\psibr$ is defined in terms of ket-vector $\psik$ \rf{07122022-manus-27} as $\langle \psi| \equiv \psik^\dagger \mubf$, where $\mubf  = e^{\irm \pi \rho_S^\vph5}$, while $\rho_S^\vph5$ is given in \rf{07122022-manus-32}. In the alternative formulation, the action and charges are obtained by using the replacement $\phi \rightarrow \psi$ in \rf{05122022-manus-01}, \rf{05122022-manus-04}.

In the CFT adapted basis, the equations of motion take simple form and therefore this basis is convenient for study of AdS/CFT correspondence for massive and partially-massless fields \cite{Metsaev:2015rda}.  For massless fields,  we have $A=0$. Therefore, for massless fields, the helicity basis and the CFT adapted basis can be used on an equal footing for study of light-cone gauge AdS/CFT correspondence (see, e.g., Refs.\cite{Skvortsov:2018uru,deMelloKoch:2022sul}). The study of AdS/CFT correspondence for arbitrary spin massive and partially-massless fields by various covariant methods may be found in Refs.\cite{Metsaev:2011uy,Bekaert:2012vt}.

%%%%%%%%%%%%%%%%%%%%%%%%%%%%%%%%%%%%%%%%%%%%%%%%%%%%%%%%%%%%%%%%%%%%%%%%%%%%%%%%%%%%%%%%%
%%%%%%%%%%%%%%%%%%%%%%%%%%%%%%%%%%%%%%%%%%%%%%%%%%%%%%%%%%%%%%%%%%%%%%%%%%%%%%%%%%%%%%%%%
\newsection{ \large Basic equations of light-cone gauge approach in $AdS_4$ }\label{sec-05}
%%%%%%%%%%%%%%%%%%%%%%%%%%%%%%%%%%%%%%%%%%%%%%%%%%%%%%%%%%%%%%%%%%%%%%%%%%%%%%%%%%%%%%%%%
%%%%%%%%%%%%%%%%%%%%%%%%%%%%%%%%%%%%%%%%%%%%%%%%%%%%%%%%%%%%%%%%%%%%%%%%%%%%%%%%%%%%%%%%%

In this section, we present basic equations of light-cone gauge approach in $AdS_4$
and briefly outline procedure of the derivation of our results in Secs.\ref{sec-03},\ref{sec-04}.%
\footnote{For $AdS_{d+1}$, $d\geq3$, the basic equations in CFT adapted basis were obtained in Ref.\cite{Metsaev:2015rda}.
}

\noinbf{Basic equations in helicity basis}. Basic equations are formulated in terms of the operators $M^{z1}$ and $B^1$, $B^z$. In terms of these operators, the operators $A$ and $B$ take the form (see Refs.\cite{Metsaev:2003cu,Metsaev:2019opn}),
\be \label{08122022-manus-01}
A  =  \CC_2 + 2B^z + 2M^{z1}M^{z1} + 2\,, \qquad  B  =  B^z + M^{z1}M^{z1}\,,
\ee
where $\CC_2$ is given in \rf{06122022-manus-09}, while the operators $B^1$, $B^z$, $M^{z1}$ satisfy the basic equations given by
\beq
\label{08122022-manus-03} && [M^{z1},B^1]= B^z\,, \qquad [M^{z1},B^z]= -B^1\,, \qquad
\\
\label{08122022-manus-04} && [B^z,B^1] = \bigl( \CC_2  +  2 M^{z1}M^{z1} + 2 \bigr)M^{z1}\,,
\eeq
and the restrictions $(\mubf B^1)^\dagger=\mubf B^1$, $(\mubf B^z)^\dagger=\mubf B^z$, $(\mubf M^{z1})^\dagger = - \mubf M^{z1}$. We also find  the following representation of the 4th-order Casimir operator of the $so(3,2)$ algebra:%
\footnote{For all solutions to spin operators obtained in this paper, we find $\CC_4 = (E_0-1)(E_0-2)s(s+1)$ as it should be.}
\be \label{24112022-04}
\CC_4 = B^1 B^1 + B^z B^z - (\CC_2 + 1 )(M^{z1})^2 - (M^{z1})^4\,.
\ee
Using helicity basis operators $B^\Rsm$, $B^\Lsm$, and $M^{\Rsm\Lsm}$ defined by the relations
\be \label{08122022-manus-05}
B^\Rsm = \frac{1}{\sqrt{2}}( B^1 + \irm B^z)\,, \qquad B^\Lsm = \frac{1}{\sqrt{2}}( B^1 - \irm B^z)\,, \qquad M^{z1} = -\irm M^{\Rsm\Lsm}\,,
\ee
we cast equations \rf{08122022-manus-03}, \rf{08122022-manus-04} into the desired helicity basis equations,
\beq
\label{08122022-manus-06} && [M^{\Rsm\Lsm}, B^\Rsm] = B^\Rsm\,, \qquad  [M^{\Rsm\Lsm}, B^\Lsm] = - B^\Lsm\,,
\\
\label{08122022-manus-07} && [B^\Rsm,B^\Lsm] = \bigl( \CC_2  -  2 M^{\Rsm\Lsm}M^{\Rsm\Lsm} + 2 \bigr)M^{\Rsm\Lsm}\,.
\eeq
In the helicity basis, the operators $A$ and $B$ \rf{08122022-manus-01} take the form
\be \label{08122022-manus-08} A  =  \CC_2 + \irm \sqrt{2}(B^\Lsm-B^\Rsm) -  2M^{\Rsm\Lsm}M^{\Rsm\Lsm} + 2\,, \qquad B  =  \frac{\irm}{\sqrt{2}}(B^\Lsm-B^\Rsm) -  M^{\Rsm\Lsm} M^{\Rsm\Lsm} \,.
\ee
We note also the restrictions $(\mubf B^\Rsm)^\dagger=\mubf B^\Lsm$, $(\mubf M^{\Rsm\Lsm})^\dagger=\mubf M^{\Rsm\Lsm}$. The $\CC_4$ \rf{24112022-04} takes the form
\be \label{24112022-06}
\CC_4 = B^\Rsm B^\Lsm + B^\Lsm B^\Rsm + (\CC_2 + 1 )(M^{\Rsm\Lsm})^2 - (M^{\Rsm\Lsm})^4\,.
\ee
Commutators given in \rf{08122022-manus-06} motivate us to look for the following solution for the operators $B^\Rsm$, $B^\Lsm$ and $M^{\Rsm\Lsm}$,
\be \label{08122022-manus-10}
B^\Rsm =  f  S^\Rsm\,, \qquad
B^\Lsm =   S^\Lsm \fb\,,\qquad M^{\Rsm\Lsm} = S\,,
\ee
where $f=f(S)$, $\fb=\fb(S)$. The $S^{\Rsm,\Lsm}$, $S$ are given in \rf{05122022-manus-19}. What is required is to find $f$ and $\fb$. Using equation \rf{08122022-manus-07}, we find the following solution for $f\fb$:
\be \label{08122022-manus-11}
f\fb = (E_0  -1 - S)(E_0  - 2 + S)\,.
\ee
To fix $f$ and $\fb$, we use $\mubf=\mubf(S)$ and note that the hermicity conditions \rf{05122022-manus-17a1} amount to the equation
\be \label{08122022-manus-12}
\mubf(S) f^\dagger(S)  = \mubf(S-1) \fb(S)\,.
\ee
Introducing the decomposition of $f\fb$ \rf{08122022-manus-11},
\be \label{08122022-manus-14}
f\fb = e^{\irm \pi \varphi_S^\vph5}|F|^2\,,
\ee
we note that equation \rf{08122022-manus-12} can be represented as
\be  \label{08122022-manus-15}
\mubf(S)  = \mubf(S-1) e^{\irm \pi \varphi_S^\vph5}\,.
\ee
For massive fields, using \rf{06122022-manus-15}, \rf{08122022-manus-11}, \rf{08122022-manus-14} , we find $\varphi_S^\vph5=0$. Therefore solution to equation \rf{08122022-manus-15} can be chosen to be $\mubf=1$. The corresponding solution for $f$, $\fb$ given in \rf{06122022-manus-08} is obtained by using equations \rf{08122022-manus-11}, \rf{08122022-manus-12}.

We now consider a partially-massless field. Using the notation $|\phi_{m,s}\rangle$ for the ket-vector of massive field in \rf{06122022-manus-01}, we  decompose the $|\phi_{m,s}\rangle$ as
\be \label{08122022-manus-16}
|\phi_{m,s}\rangle  = |\phi_\msv\rangle + |\phi^\opsm\rangle + |\phi^\omsm\rangle\,, \qquad \quad
|\phi_\msv\rangle \equiv \sum_{n\in [-s_\msv,s_\msv]} \phi_n(x,z) |n\rangle\,,
\ee
where $s_\msv \equiv s-1-t$, while the ket-vectors $|\phi^{\opsm,\omsm}\rangle$ are defined in \rf{06122022-manus-36}. Plugging formally $E_0$ corresponding to partially-massless field \rf{06122022-manus-35} into expressions for massive field in \rf{06122022-manus-07}-\rf{06122022-manus-09}, we verify that the ket-vectors $|\phi_\msv\rangle$, $|\phi^{\opsm,\omsm}\rangle$ form invariant subspaces under action of the operators $A$, $B$, and $M^{z1}$ and the action for massive field \rf{05122022-manus-01} is then decomposed as
\be \label{08122022-manus-17}
S(\phi_{m,s}) = S(\phi_\msv) + S(\phi^\opsm) + S(\phi^\omsm)\,.
\ee
Note that the Noether charges \rf{05122022-manus-04}  are also decomposed as in \rf{08122022-manus-17}. This is to say that, for $E_0$ given in \rf{06122022-manus-35}, the ket-vector $|\phi_{m,s}\rangle$ is decomposed into three decoupled systems - one massive spin-$s_\msv$ field $|\phi_\msv\rangle$, one helicity $\lambda=s$ and depth-$t$ partially-massless field $|\phi^\opsm\rangle$ and one helicity $\lambda=-s$ and depth-$t$ partially-massless field $|\phi^\omsm\rangle$. However plugging $E_0$  \rf{06122022-manus-35} into \rf{06122022-manus-07}-\rf{06122022-manus-09}, we get non-hermitian actions for the partially-massless fields \rf{08122022-manus-17}. Hermitian actions and the Noether charges for $\phi^{\opsm,\omsm}$ are obtained by choosing a suitable $\mubf$. Namely, plugging $E_0$ \rf{06122022-manus-35} into \rf{08122022-manus-11} and using \rf{08122022-manus-14}, we find $\varphi_S^\vph5=1$
This implies that, for the partially-massless fields, equation \rf{08122022-manus-15} takes the form $\mubf(S)=-\mubf(S-1)$. Solutions to such equation can be chosen as in \rf{06122022-manus-39}. The corresponding solutions for $f$, $\fb$ given in \rf{06122022-manus-38} are found by using equations \rf{08122022-manus-11}, \rf{08122022-manus-12}.

\noinbf{ Basic equations in CFT adapted basis}. The basic equations are now formulated in terms of operators $\nu$, $W^1$, $\Wb^1$. In terms of these operators, the operators $A$, $B$, and $M^{z1}$ are expressed as
\be
\label{08122022-manus-18}  A  = \nu^2 - \frac{1}{4}\,, \qquad B = \half \Bigl( \nu^2  - \CC_2   - \frac{9}{4} \Bigr)\,, \qquad M^{z1} = W^1 - \Wb^1\,,
\ee
where $\CC_2$ is given in \rf{06122022-manus-09}. The basic equations for the operators $\nu$, $W^1$, $\Wb^1$ take then the form
\beq
\label{08122022-manus-20} && [\nu, W^1] = W^1\,,\qquad [\nu,\Wb^1]= - \Wb^1\,,
\\
\label{08122022-manus-21} && 2(\nu-1) W^1 \Wb^1 - 2(\nu+1) \Wb^1 W^1   = B\,.
\eeq
We note also the restrictions $(\mubf \nu)^\dagger=\mubf \nu$, $(\mubf W^1)^\dagger=\mubf \Wb^1$.
In the CFT adapted basis, the 4th-order Casimir operator \rf{24112022-04}  is represented as
\be
\CC_4 =   \frac{1}{4} \bigl( \nu^2 - \CC_2 - \frac{9}{4} \bigr)^2 + (2\nu+1)(\nu-1) W^1 \Wb^1  + (2\nu-1)(\nu+1)  \Wb^1 W^1\,.
\ee
General results in Ref.\cite{Metsaev:2015rda} suggest the following representation for the operators $W^1$, $\Wb^1$ and $\nu$,
\be \label{08122022-manus-22}
\nu = \kappa  + S\,, \qquad  W^1 = f  S^\Rsm\,,\qquad \Wb^1 =   S^\Lsm \fb \,,
\ee
where $f=f(S)$, $\fb=\fb(S)$, while $\kappa$ is given in \rf{07122022-manus-15}. Equation \rf{08122022-manus-21} leads to the solution for  $f\fb$:
\be \label{08122022-manus-23}
f\fb  =  \frac{(2\kappa  +  s+S)(2\kappa -1 - s + S)}{8(\kappa -1+S)(\kappa  + S)}\,.
\ee
To fix $f$ and $\fb$, we use equations \rf{08122022-manus-12}-\rf{08122022-manus-15} and \rf{08122022-manus-23}. For massive fields, using \rf{06122022-manus-15}, \rf{08122022-manus-14}, \rf{08122022-manus-23}, we find $\varphi=0$. Using  equation \rf{08122022-manus-15}, we find then the solution $\mubf=1$ , while the corresponding $f$, $\fb$ given in \rf{07122022-manus-18} are obtained by using equations \rf{08122022-manus-12}, \rf{08122022-manus-23}.

Now consider partially-massless field. Using the notation $|\phi_{m,s}\rangle$ for ket-vector of massive field in \rf{06122022-manus-01} and the notation $|\phi_\pms\rangle$ for ket-vector \rf{07122022-manus-19}, we  decompose the $|\phi_{m,s}\rangle$ as
\be \label{08122022-manus-24}
|\phi_{m,s}\rangle  = |\phi_\msv\rangle + |\phi_\pms\rangle\,, \qquad \quad
|\phi_\msv\rangle \equiv \sum_{n\in [-s+2+2t,s]} \phi_n(x,z) |n\rangle\,.
\ee
Plugging $E_0$ corresponding to partially-massless field \rf{06122022-manus-35} into expressions for massive field in \rf{07122022-manus-15}, \rf{07122022-manus-18}, we verify that the ket-vectors $|\phi_\msv\rangle$, $|\phi_\pms\rangle$ \rf{08122022-manus-24} form invariant subspaces under action of operators $A$, $B$, and $M^{z1}$ and the Noether charges for massive field \rf{05122022-manus-04} are decomposed as%
\footnote{In the CFT adapted basis, the operator $A$ is diagonal. Therefore, in the CFT adapted basis, the decomposition for the action of massive field $S(\phi_{m,s}) = S(\phi_\msv) + S(\phi_\pms)$ holds true for arbitrary $E_0$.
}
\be \label{08122022-manus-25}
G_\field(\phi_{m,s}) = G_\field(\phi_\msv) + G_\field(\phi_\pms)\,.
\ee
In other words, for $E_0$ \rf{06122022-manus-35}, the ket-vector $|\phi_{m,s}\rangle$ is decomposed into two decoupled systems - one massive spin-$s_\msv$ field $|\phi_\msv\rangle$, where $s_\msv=s-1-t$, and one spin-$s$ and depth-$t$ partially-massless field $|\phi_\pms\rangle$. Note however that plugging $E_0$  \rf{06122022-manus-35} into \rf{07122022-manus-15}, \rf{07122022-manus-18}, we get non-hermitian Noether charges for partially-massless field  \rf{05122022-manus-04}. Hermitian Noether charges are obtained by choosing a suitable $\mubf$. Namely, plugging $E_0$ \rf{06122022-manus-35} into \rf{08122022-manus-23} and using \rf{08122022-manus-14}, we find $\varphi_S^\vph5=\rho_S^\vph5+\rho_{S-1}^\vph5$, where $\rho_S^\vph5$ is given in \rf{07122022-manus-20}.
Using such $\varphi_S^\vph5$ and equation \rf{08122022-manus-15}, we find $\mubf  = \exp(\irm \pi \rho_S^\vph5)$, while rhe expressions for $f$, $\fb$ given in \rf{06122022-manus-38} are found by using equations \rf{08122022-manus-12}, \rf{08122022-manus-23}.

\medskip

\noindent {\bf Conclusions}. In this paper, we applied light-cone gauge approach for the study of Lagrangian formulation of massive and partially-massless fields in $AdS_4$. We studied both the bosonic and fermionic fields. In our approach, we used bosonic spinor-like oscillators. In Ref.\cite{Metsaev:2022yvb}, we demonstrated that the use of the bosonic spinor-like oscillators allows us to find the simple solution for all cubic vertices for massive fields in flat space, while, in Ref.\cite{Metsaev:2018xip}, we developed the method for study of interacting fields in $AdS_4$.
We expect therefore that the results in this paper as well as the methods in   Refs.\cite{Metsaev:2022yvb,Metsaev:2018xip} will provide us new interesting possibilities for building interaction vertices of the massive and partially-massless fields in $AdS_4$.
For the extensive study of interacting partially-massless fields, see Ref.\cite{Basile:2022mif} and Refs.\cite{Joung:2012rv}-\cite{Grigoriev:2020lzu}.
We note that the formalism we used in this paper seems to be closely related to twistor approach. Recent interesting application of twistor approach for study of higher-spin fields may be found in Refs.\cite{Krasnov:2021nsq}-\cite{Tran:2022tft}.
Supermultiplets of partially-massless fields were studied in Refs.\cite{Garcia-Saenz:2018wnw}-\cite{Hutchings:2021moq}. We believe that the light-cone gauge formalism in this paper and the method for the study of supersymmetric theories in Refs.\cite{Metsaev:2019dqt,Metsaev:2019aig} will provide us new possibilities for investigation of interacting partially-massless supermultiplets.
We expect that our result will also be helpful for study of quantum massive and partially-massless fields. Application of light-cone approach for the study of quantum corrections in chiral higher-spin gravity may be found in Refs.\cite{Skvortsov:2018jea}. Computation of quantum corrections in higher-spin gravity by using the covariant de Donder gauge may be found in Ref.\cite{Ponomarev:2016jqk}. Study of partially-massless fields along the lines in Refs.\cite{Najafizadeh:2018cpu} could also be of some interest.

%{\bf Acknowledgments}. This work was supported by the RFBR Grant No.20-02-00193.

\vspace{-0.3cm}
%%%%%%%%%%%%%%%%%%%%%%%%%%%%%%%%%%%%%%%%%%%%%%%%%%%%%%%%%%%%%%%%%%%%%%
\setcounter{section}{0} \setcounter{subsection}{0}
%%%%%%%%%%%%%%%%%%%%%%%%%%%%%%%%%%%%%%%%%%%%%%%%%%%%%%%%%%%%%%%%%%%%%%


\small
\begin{thebibliography}{30}

\parskip=-6pt


%\cite{Brink:2005wh}
\bibitem{Brink:2005wh}
L.~Brink,
``Particle physics as representations of the Poincare algebra,''
Lecture presented at the Poincar´e Symposium held in Brussels on October 8-9, 2004. [arXiv:hep-th/0503035 [hep-th]].



%\cite{Green:1983hw}
\bibitem{Green:1983hw}
L.~Brink, M.~B.~Green and J.~H.~Schwarz,
%``Superfield Theory of Type II Superstrings,''
Nucl. Phys. B \textbf{219} (1983), 437-478
%doi:10.1016/0550-3213(83)90651-X



%\cite{Mandelstam:1982cb}
\bibitem{Mandelstam:1982cb}
S.~Mandelstam,
%``Light Cone Superspace and the Ultraviolet Finiteness of the N=4 Model,''
Nucl. Phys. B \textbf{213} (1983), 149-168
%doi:10.1016/0550-3213(83)90179-7


%\cite{Brink:1982wv}
\bibitem{Brink:1982wv}
L.~Brink, O.~Lindgren and B.~E.~W.~Nilsson,
%``The Ultraviolet Finiteness of the N=4 Yang-Mills Theory,''
Phys. Lett. B \textbf{123} (1983), 323-328
%doi:10.1016/0370-2693(83)91210-8




%\cite{Bengtsson:1983pd}
\bibitem{Bengtsson:1983pd}
A.~K.~H.~Bengtsson, I.~Bengtsson and L.~Brink,
%``Cubic Interaction Terms for Arbitrary Spin,''
Nucl. Phys. B \textbf{227} (1983), 31-40
%doi:10.1016/0550-3213(83)90140-2


\bibitem{Bengtsson:1983pg}
A.~K.~H.~Bengtsson, I.~Bengtsson and L.~Brink,
%``Cubic Interaction Terms for Arbitrarily Extended Supermultiplets,''
Nucl. Phys. B \textbf{227} (1983), 41-49
%doi:10.1016/0550-3213(83)90141-4



%\cite{Bengtsson:2012dw}
\bibitem{Bengtsson:2012dw}
A.~K.~H.~Bengtsson, L.~Brink and S.~S.~Kim,
%``Counterterms in Gravity in the Light-Front Formulation and a D=2 Conformal-like Symmetry in Gravity,''
JHEP \textbf{03} (2013), 118
%doi:10.1007/JHEP03(2013)118
[arXiv:1212.2776 [hep-th]].


%\cite{Metsaev:1999ui}
\bibitem{Metsaev:1999ui}
  R.~R.~Metsaev,
  %``Light cone form of field dynamics in Anti-de Sitter space-time and AdS / CFT correspondence,''
  Nucl.\ Phys.\ B {\bf 563}, 295 (1999)
%  doi:10.1016/S0550-3213(99)00554-4
  [hep-th/9906217].
  %%CITATION = doi:10.1016/S0550-3213(99)00554-4;%%



%\cite{Metsaev:2003cu}
\bibitem{Metsaev:2003cu}
  R.~R.~Metsaev,
  %``Massive totally symmetric fields in AdS(d),''
  Phys.\ Lett.\ B {\bf 590}, 95 (2004)
%  doi:10.1016/j.physletb.2004.03.057
  [hep-th/0312297].






%\cite{Zinoviev:2001dt}
\bibitem{Zinoviev:2001dt}
Y.~M.~Zinoviev,
%``On massive high spin particles in AdS,''
[arXiv:hep-th/0108192 [hep-th]].

%\cite{Metsaev:2006zy}
\bibitem{Metsaev:2006zy}
R.~R.~Metsaev,
%``Gauge invariant formulation of massive totally symmetric fermionic fields in (A)dS space,''
Phys. Lett. B \textbf{643} (2006), 205-212
%doi:10.1016/j.physletb.2006.11.002
[arXiv:hep-th/0609029 [hep-th]].



%\cite{Zinoviev:2008ze}
\bibitem{Zinoviev:2008ze}
Y.~M.~Zinoviev,
%``Frame-like gauge invariant formulation for massive high spin particles,''
Nucl. Phys. B \textbf{808} (2009), 185-204
%doi:10.1016/j.nuclphysb.2008.09.020
[arXiv:0808.1778 [hep-th]].

%\cite{Ponomarev:2010st}
\bibitem{Ponomarev:2010st}
D.~S.~Ponomarev and M.~A.~Vasiliev,
%``Frame-Like Action and Unfolded Formulation for Massive Higher-Spin Fields,''
Nucl. Phys. B \textbf{839} (2010), 466-498
%doi:10.1016/j.nuclphysb.2010.06.007
[arXiv:1001.0062 [hep-th]].



%\cite{Skvortsov:2006at}
\bibitem{Skvortsov:2006at}
E.D. Skvortsov and M.A. Vasiliev,
%``Geometric formulation for partially massless fields,''
Nucl. Phys. B \textbf{756} (2006), 117-147
%doi:10.1016/j.nuclphysb.2006.06.019
[arXiv:hep-th/0601095].


%\cite{Khabarov:2020glf}
\bibitem{Khabarov:2020glf}
M.~V.~Khabarov and Y.~M.~Zinoviev,
%``Massive higher spin supermultiplets unfolded,''
Nucl. Phys. B \textbf{953} (2020), 114959
%doi:10.1016/j.nuclphysb.2020.114959
[arXiv:2001.07903 [hep-th]].


%\cite{Khabarov:2019dvi}
\bibitem{Khabarov:2019dvi}
M.~V.~Khabarov and Y.~M.~Zinoviev,
%``Massive higher spin fields in the frame-like multispinor formalism,''
Nucl. Phys. B \textbf{948} (2019), 114773
%doi:10.1016/j.nuclphysb.2019.114773
[arXiv:1906.03438 [hep-th]].



%\cite{Buchbinder:2006ge}
\bibitem{Buchbinder:2006ge}
I.~L.~Buchbinder, V. Krykhtin, P. Lavrov,
%``Gauge invariant Lagrangian formulation of higher spin massive bosonic field theory in AdS space,''
Nucl. Phys. B \textbf{762} (2007), 344-376
%doi:10.1016/j.nuclphysb.2006.11.021
[arXiv:hep-th/0608005].
%
\\
%
%\cite{Buchbinder:2007vq}
%\bibitem{Buchbinder:2007vq}
I.~L.~Buchbinder, V. Krykhtin, A. Reshetnyak,
%``BRST approach to Lagrangian construction for fermionic higher spin fields in (A)dS space,''
Nucl. Phys. B \textbf{787} (2007), 211-240
%doi:10.1016/j.nuclphysb.2007.06.006
[hep-th/0703049].




%\cite{Alkalaev:2009vm}
\bibitem{Alkalaev:2009vm}
K.~B.~Alkalaev and M.~Grigoriev,
%``Unified BRST description of AdS gauge fields,''
Nucl. Phys. B \textbf{835} (2010), 197-220
%doi:10.1016/j.nuclphysb.2010.04.004
[arXiv:0910.2690 [hep-th]].
%
\\
%
%\cite{Alkalaev:2011zv}
%\bibitem{Alkalaev:2011zv}
K.~Alkalaev and M.~Grigoriev,
%``Unified BRST approach to (partially) massless and massive AdS fields of arbitrary symmetry type,''
Nucl. Phys. B \textbf{853} (2011), 663-687
%doi:10.1016/j.nuclphysb.2011.08.005
[arXiv:1105.6111 [hep-th]].



%\cite{Deser:1983mm}
\bibitem{Deser:1983mm}
S.~Deser and R.~I.~Nepomechie,
%``Gauge Invariance Versus Masslessness In De Sitter Space,''
Annals Phys.\  {\bf 154}, 396 (1984).
%%CITATION = APNYA,154,396;%%


%\cite{Deser:2001xr}
\bibitem{Deser:2001xr}
S.~Deser and A.~Waldron,
%``Null propagation of partially massless higher spins in (A)dS and cosmological constant speculations,''
Phys. Lett. B \textbf{513} (2001), 137-141
%doi:10.1016/S0370-2693(01)00756-0
[arXiv:hep-th/0105181 [hep-th]].



%\cite{Basile:2022mif}
\bibitem{Basile:2022mif}
T.~Basile, S.~Dhasmana and E.~Skvortsov,
%``Chiral approach to partially-massless fields,''
[arXiv:2212.06226 [hep-th]].






%\cite{Metsaev:2019opn}
\bibitem{Metsaev:2019opn}
R.~R.~Metsaev,
%``Light-cone continuous-spin field in AdS space,''
Phys. Lett. B \textbf{793} (2019), 134-140
%doi:10.1016/j.physletb.2019.04.041
[arXiv:1903.10495 [hep-th]].
%
\\
%
%\cite{Metsaev:2021zdg}
%\bibitem{Metsaev:2021zdg}
R.~R.~Metsaev,
%``Mixed-symmetry continuous-spin fields in flat and AdS spaces,''
Phys. Lett. B \textbf{820} (2021), 136497
%doi:10.1016/j.physletb.2021.136497
[arXiv:2105.11281 [hep-th]].



%\cite{Metsaev:2015rda}
\bibitem{Metsaev:2015rda}
R.~R.~Metsaev,
%``Light-cone AdS/CFT-adapted approach to AdS fields/currents, shadows, and conformal fields,''
JHEP \textbf{10} (2015), 110
%doi:10.1007/JHEP10(2015)110
[arXiv:1507.06584 [hep-th]].


%\cite{Brink:2000ag}
\bibitem{Brink:2000ag}
L.~Brink, R.~R.~Metsaev, M.~A.~Vasiliev,
%``How massless are massless fields in AdS(d),''
Nucl. Phys. B \textbf{586} (2000), 183-205
%doi:10.1016/S0550-3213(00)00402-8
[arXiv:hep-th/0005136]





\bibitem{Evans}
N.~T.~Evans,
%``Discrete Series for the Universal Covering Group of the 3+3
%dimensional de Sitter Group''
J.\ Math.\ Phys.\ {\bf 8},  170 (1967).


%\cite{Metsaev:1995re}
\bibitem{Metsaev:1995re}
R.~R.~Metsaev,
%``Massless mixed symmetry bosonic free fields in d-dimensional anti-de Sitter space-time,''
Phys. Lett. B \textbf{354} (1995), 78-84
%doi:10.1016/0370-2693(95)00563-Z


%\cite{Ponomarev:2022vjb}
\bibitem{Ponomarev:2022vjb}
D.~Ponomarev,
``Basic introduction to higher-spin theories,''
[arXiv:2206.15385 [hep-th]].





%\cite{Skvortsov:2018uru}
\bibitem{Skvortsov:2018uru}
  E.~Skvortsov,
  %``Light-Front Bootstrap for Chern-Simons Matter Theories,''
  JHEP {\bf 1906}, 058 (2019)
% doi:10.1007/JHEP06(2019)058
  [arXiv:1811.12333 [hep-th]].
  %%CITATION = doi:10.1007/JHEP06(2019)058;%%

%\cite{deMelloKoch:2022sul}
\bibitem{deMelloKoch:2022sul}
R.~de Mello Koch and G.~Kemp,
%``Holography of Information in AdS/CFT,''
[arXiv:2210.11066 [hep-th]].

%\cite{Metsaev:2011uy}
\bibitem{Metsaev:2011uy}
R.~R.~Metsaev,
%``Anomalous conformal currents, shadow fields and massive AdS fields,''
Phys. Rev. D \textbf{85} (2012), 126011
%doi:10.1103/PhysRevD.85.126011
[arXiv:1110.3749 [hep-th]].


%\cite{Bekaert:2012vt}
\bibitem{Bekaert:2012vt}
X.~Bekaert and M.~Grigoriev,
%``Notes on the ambient approach to boundary values of AdS gauge fields,''
J. Phys. A \textbf{46} (2013), 214008
%doi:10.1088/1751-8113/46/21/214008
[arXiv:1207.3439 [hep-th]].
%
\\
%
%\cite{Bekaert:2013zya}
%\bibitem{Bekaert:2013zya}
X.~Bekaert and M.~Grigoriev,
%``Higher order singletons, partially massless fields and their boundary values in the ambient approach,''
Nucl. Phys. B \textbf{876} (2013), 667-714
%doi:10.1016/j.nuclphysb.2013.08.015
[arXiv:1305.0162 [hep-th]].


%\cite{Metsaev:2022yvb}
\bibitem{Metsaev:2022yvb}
R.~R.~Metsaev,
%``Interacting massive and massless arbitrary spin fields in 4d flat space,''
Nucl. Phys. B \textbf{984} (2022), 115978
%doi:10.1016/j.nuclphysb.2022.115978
[arXiv:2206.13268 [hep-th]].


%\cite{Metsaev:2018xip}
\bibitem{Metsaev:2018xip}
  R.~R.~Metsaev,
  %``Light-cone gauge cubic interaction vertices for massless fields in AdS(4),''
  Nucl.\ Phys.\ B {\bf 936}, 320 (2018)
%  doi:10.1016/j.nuclphysb.2018.09.021
  [arXiv:1807.07542 [hep-th]].




%\cite{Joung:2012rv}
\bibitem{Joung:2012rv}
E.~Joung, L.~Lopez and M.~Taronna,
%``On the cubic interactions of massive and partially-massless higher spins in (A)dS,''
JHEP \textbf{07} (2012), 041
%doi:10.1007/JHEP07(2012)041
[arXiv:1203.6578 [hep-th]].

%\cite{Zinoviev:2014zka}
\bibitem{Zinoviev:2014zka}
Y.~M.~Zinoviev,
%``Massive spin-2 in the Fradkin\textendash{}Vasiliev formalism. I. Partially massless case,''
Nucl. Phys. B \textbf{886} (2014), 712-732
%doi:10.1016/j.nuclphysb.2014.07.013
[arXiv:1405.4065 [hep-th]].

%\cite{Joung:2019wwf}
\bibitem{Joung:2019wwf}
E.~Joung, K.~Mkrtchyan and G.~Poghosyan,
%``Looking for partially-massless gravity,''
JHEP \textbf{07} (2019), 116
%doi:10.1007/JHEP07(2019)116
[arXiv:1904.05915 [hep-th]].


%\cite{Boulanger:2019zic}
\bibitem{Boulanger:2019zic}
N.Boulanger, C.Deffayet, S.Garcia-Saenz, L.Traina,
%``Theory for multiple partially massless spin-2 fields,''
Phys.Rev.D100 (2019) 101701
%doi:10.1103/PhysRevD.100.101701
arXiv:1906.03868



%\cite{Grigoriev:2020lzu}
\bibitem{Grigoriev:2020lzu}
  M.~Grigoriev, K.~Mkrtchyan, E.~Skvortsov,
  %``Matter-free higher spin gravities in 3D: Partially-massless fields and general structure,''
  Phys.Rev.D {\bf 102}, no. 6, 066003 (2020)
%  doi:10.1103/PhysRevD.102.066003
  [arXiv:2005.05931].
  %%CITATION = doi:10.1103/PhysRevD.102.066003;%%

%\cite{Krasnov:2021nsq}
\bibitem{Krasnov:2021nsq}
K.~Krasnov, E.~Skvortsov and T.~Tran,
%``Actions for self-dual Higher Spin Gravities,''
JHEP \textbf{08} (2021), 076
%doi:10.1007/JHEP08(2021)076
[arXiv:2105.12782 [hep-th]].

%\cite{Adamo:2022lah}
\bibitem{Adamo:2022lah}
T.~Adamo and T.~Tran,
%``Higher-spin Yang-Mills, amplitudes and self-duality,''
[arXiv:2210.07130 [hep-th]].

%\cite{Steinacker:2022jjv}
\bibitem{Steinacker:2022jjv}
H.~C.~Steinacker and T.~Tran,
%``A twistorial description of the IKKT-matrix model,''
JHEP \textbf{11} (2022), 146
%doi:10.1007/JHEP11(2022)146
[arXiv:2203.05436 [hep-th]].

%\cite{Tran:2022tft}
\bibitem{Tran:2022tft}
T.~Tran,
%``Toward a twistor action for chiral higher-spin gravity,''
[arXiv:2209.00925 [hep-th]].


%\cite{Garcia-Saenz:2018wnw}
\bibitem{Garcia-Saenz:2018wnw}
S.~Garcia-Saenz, K.~Hinterbichler and R.~A.~Rosen,
%``Supersymmetric Partially Massless Fields and Non-Unitary Superconformal Representations,''
JHEP \textbf{11} (2018), 166
%doi:10.1007/JHEP11(2018)166
[arXiv:1810.01881 [hep-th]].


%\cite{Bittermann:2020xkl}
\bibitem{Bittermann:2020xkl}
N.Bittermann, S.Garcia-Saenz, K.Hinterbichler, R.Rosen,
%``$ \mathcal{N} $ = 2 supersymmetric partially massless fields and other exotic non-unitary superconformal representations,''
JHEP \textbf{08} (2021), 115
%doi:10.1007/JHEP08(2021)115
[arXiv:2011.05994]

%\cite{Buchbinder:2019olk}
\bibitem{Buchbinder:2019olk}
I.~L.~Buchbinder, M. Khabarov, T. Snegirev, Y.M. Zinoviev,
%``Lagrangian description of the partially massless higher spin N = 1 supermultiplets in AdS$_{4}$ space,''
JHEP \textbf{08} (2019), 116
%doi:10.1007/JHEP08(2019)116
[arXiv:1904.01959].


%\cite{Buchbinder:2021oiw}
\bibitem{Buchbinder:2021oiw}
E.~I.~Buchbinder, D.~Hutchings, S.~M.~Kuzenko, M.~Ponds,
%``AdS superprojectors,''
JHEP \textbf{04} (2021), 074
%doi:10.1007/JHEP04(2021)074
[arXiv:2101.05524].


%\cite{Hutchings:2021moq}
\bibitem{Hutchings:2021moq}
D.~Hutchings, S.~M.~Kuzenko and M.~Ponds,
%``AdS (super)projectors in three dimensions and partial masslessness,''
JHEP \textbf{10} (2021), 090
%doi:10.1007/JHEP10(2021)090
[arXiv:2107.12201 [hep-th]].

%\cite{Metsaev:2019dqt}
\bibitem{Metsaev:2019dqt}
  R.~R.~Metsaev,
  %``Cubic interaction vertices for N=1 arbitrary spin massless supermultiplets in flat space,''
  JHEP {\bf 1908}, 130 (2019)
%  doi:10.1007/JHEP08(2019)130
  [arXiv:1905.11357 [hep-th]].
  %%CITATION = doi:10.1007/JHEP08(2019)130;%%

%\cite{Metsaev:2019aig}
\bibitem{Metsaev:2019aig}
  R.~R.~Metsaev,
  %``Cubic interactions for arbitrary spin $ \mathcal{N} $ -extended massless supermultiplets in 4d flat space,''
  JHEP {\bf 1911}, 084 (2019)
%  doi:10.1007/JHEP11(2019)084
  [arXiv:1909.05241 [hep-th]].
  %%CITATION = doi:10.1007/JHEP11(2019)084;%%


%\cite{Skvortsov:2018jea}
\bibitem{Skvortsov:2018jea}
E.~D.~Skvortsov, T.~Tran and M.~Tsulaia,
%``Quantum Chiral Higher Spin Gravity,''
Phys. Rev. Lett. \textbf{121} (2018) no.3, 031601
%doi:10.1103/PhysRevLett.121.031601
[arXiv:1805.00048]
%
\\
%
%\cite{Skvortsov:2020wtf}
%\bibitem{Skvortsov:2020wtf}
E.~Skvortsov, T.~Tran and M.~Tsulaia,
%``More on Quantum Chiral Higher Spin Gravity,''
Phys. Rev. D \textbf{101} (2020) no.10, 106001
%doi:10.1103/PhysRevD.101.106001
[arXiv:2002.08487]
%
\\
%
%\cite{Tsulaia:2022csz}
%\bibitem{Tsulaia:2022csz}
M.~Tsulaia and D.~Weissman,
%``Supersymmetric quantum chiral higher spin gravity,''
JHEP \textbf{12} (2022), 002
%doi:10.1007/JHEP12(2022)002
[arXiv:2209.13907]


%\cite{Ponomarev:2016jqk}
\bibitem{Ponomarev:2016jqk}
D.~Ponomarev and A.~A.~Tseytlin,
%``On quantum corrections in higher-spin theory in flat space,''
JHEP \textbf{05} (2016), 184
%doi:10.1007/JHEP05(2016)184
[arXiv:1603.06273 [hep-th]].



%\cite{Najafizadeh:2018cpu}
\bibitem{Najafizadeh:2018cpu}
M.~Najafizadeh,
%``Local action for fermionic unconstrained higher spin gauge fields in AdS and dS spacetimes,''
Phys. Rev. D \textbf{98} (2018) no.12, 125012
%doi:10.1103/PhysRevD.98.125012
[arXiv:1807.01124 [hep-th]].
%
\\
%
%\cite{Najafizadeh:2020moz}
%\bibitem{Najafizadeh:2020moz}
M.~Najafizadeh,
%``Unconstrained massless higher spin supermultiplet in AdS4 spacetime,''
Phys. Rev. D \textbf{105} (2022) no.2, 025001
%doi:10.1103/PhysRevD.105.025001
[arXiv:2012.15580 [hep-th]].







\end{thebibliography}
\end{document}